\documentclass[twocolumn]{aastex631}
\usepackage{amsmath}
\usepackage{xspace}

\newcommand{\xmm}{{\it XMM-Newton}\xspace}
\newcommand{\chandra}{{\it Chandra}\xspace}

\newcommand{\target}{AT2019qiz\xspace}

\expandafter\ifx\csname natexlab\endcsname\relax\fi
\providecommand{\url}[1]{\href{#1}{#1}}
\providecommand{\dodoi}[1]{doi:~\href{http://doi.org/#1}{\nolinkurl{#1}}}
\providecommand{\doeprint}[1]{\href{http://ascl.net/#1}{\nolinkurl{http://ascl.net/#1}}}
\providecommand{\doarXiv}[1]{\href{https://arxiv.org/abs/#1}{\nolinkurl{https://arxiv.org/abs/#1}}}

\shorttitle{Identifying QPEs with rising IR Dust Echoes}
\shortauthors{Pasham et al.}

\graphicspath{{./}{figures/}}

\begin{document}


\title{Using Infrared Dust Echoes to Identify Bright Quasi-periodic Eruption Sources}

	\author[0000-0003-1386-7861]{Dheeraj Pasham}
	\affiliation{MIT Kavli Institute for Astrophysics and Space Research \\
		Cambridge, MA 02139, USA}

\author[0000-0003-3765-6401]{E.~R.~Coughlin}
\affiliation{Department of Physics, Syracuse University, Syracuse, NY 13210, USA}

\author{S.~van Velzen}
\affiliation{Leiden Observatory, Leiden University,  Postbus 9513, 2300 RA Leiden, The Netherlands}

\author{J. Hinkle}
\affiliation{Department of Astronomy, University of Hawaii}

\begin{abstract}
Quasi-periodic eruptions (QPEs) are recurring soft X-ray outbursts from galactic nuclei and represent an intriguing new class of transients. Currently, 10 QPE sources are reported in the literature, and a major challenge lies in identifying more because they are (apparently) intrinsically and exclusively X-ray bright. Here we highlight the unusual infrared (IR) echo of the tidal disruption event (TDE) -- and subsequent QPE source -- AT2019qiz, which rose continuously and approximately linearly with time over roughly 1000 days (between 2019 and 2024). We argue that this continuous long rise alongside the relatively high inferred IR temperature (800-1200 K) cannot be generated by the TDE itself, including the late-time/remnant TDE disk, but that the reprocessing of the light from the QPEs by a shell of dust can reproduce the observations. This model predicts 1) IR QPEs at the 0.1 percent level that are potentially detectable with the James Webb Space Telescope, and 2) that if the QPEs cease in AT2019qiz, the IR light curve should decline steadily and linearly over the same 1000-day timescale. We identify another TDE with similar IR behavior, AT2020ysg, which could thus harbor QPEs. Our findings and inferences constitute a novel method for identifying ``bright'' QPEs (with peak bolometric luminosities $\gtrsim$10$^{44}$ erg/sec), i.e., that the follow-up of optically selected TDEs with wide-field infrared surveys can indirectly reveal the presence of QPEs. This approach could be particularly effective with the upcoming Roman telescope, which could detect dozens of QPE candidates for high-cadence X-ray follow-up. 


\end{abstract}



\section{Introduction}

Quasi-periodic eruptions (QPEs) 
are repeating soft X-ray (0.2-2.0 keV) signals observed from the centers of galaxies \citep[e.g.,][]{gsndisc, qpe12, qpe34}. Alongside other 
repeating extragalactic nuclear transients (RENTs) found in X-rays -- such as quasi-periodic oscillations (QPOs; \citealt{2013ApJ...776L..10L, 2019Sci...363..531P, 2025arXiv250101581M}), quasi-periodic outflows (QPOuts; \citealt{2024SciA...10J8898P}), and some repeating tidal disruption events (TDEs; e.g., \citealt{2023ApJ...942L..33W, 2023A&A...669A..75L, 2024ApJ...971L..31P, 2024A&A...683L..13L}) -- these phenomena may indicate the presence of an orbiter about a supermassive black hole, i.e., extreme mass ratio binaries. Some of these binaries might also emit gravitational waves detectable by LISA and TianQin in the next decade \citep[e.g.,][]{2022ApJ...930..122C, 2024MNRAS.527.2756Y, 2024MNRAS.532.2143K}. 

Repeating TDEs can coincide with the partial disruption of a star, and 
three out of the four known soft X-ray QPOs, as well as the only QPOuts source ASASSN-20qc, are linked to nuclear outbursts presumably fueled by a TDE. 
A connection between optically selected TDEs and QPEs is also beginning to emerge \citep[][]{2021ApJ...921L..40C, 2023A&A...675A.152Q, 2019qiz, bykov}, and \target is the first TDE to show clear QPEs four years after the initial disruption \citep{2019qiz}, thereby firmly linking the two phenomena. 

Given this connection, one method for finding more than the $\sim 10$ currently known QPEs 
would be to follow up every TDE with high-cadence X-ray observations. 
However, this method is expensive and infeasible, especially with an expected TDE discovery rate of 100s per year by Rubin observatory \citep[e.g.,][]{2020ApJ...890...73B}. The fraction of TDEs that host QPEs is also not known, meaning that such a blind follow-up could yield relatively few detections. An additional and indirect (i.e., non-X-ray) observational feature of TDEs (or AGN activity in general), suggestive of the presence of QPEs, is therefore needed to narrow the number of potential follow-up targets and increase the QPE detection probability. 

Because QPEs represent a $\sim$ constant-luminosity (averaged over the QPE recurrence time) source, one possibility for this additional feature is reprocessed dust emission. We therefore analyzed the long-term infrared light curves (from {\it NEOWISE}) of \target and all other ZTF-I TDEs \citep{2021ApJ...908....4V, 2023ApJ...942....9H}, which represents a uniform TDE sample discovered by the Zwicky Transient Facility (ZTF; \citealt{ZTF}) from 2018 - 2020. Each source has a similar infrared observational baseline of approximately 1500 days after their optical discovery and several years before the tidal disruption event, with one infrared measurement taken every six months. Our main finding is that \target displays an atypical, rising IR light curve that plateaus around 1000 days, suggestive of 
an active and dust-illuminating source. This source could be the TDE itself, a remnant TDE disk, and/or QPEs. Our analysis suggests that the first two are not consistent with the data, while the QPEs can uniquely explain both the rise time of the dust echo and the relatively high dust temperature. 

We present the data in Section \ref{sec:data}, discuss our dust model in Section \ref{sec:model}, and the predictions in Section \ref{sec:summ}.


\section{Data}\label{sec:data}
We use a 
$\Lambda$CDM cosmology with parameters H$_{0}$ = 67.4 km s$^{-1}$ Mpc$^{-1}$, $\Omega_{\rm m}$ = 0.315 and $\Omega_{\rm \Lambda}$ = 1 - $\Omega_{\rm m}$ = 0.685 \citep{2020A&A...641A...6P}. Using the Cosmology calculator of \cite{2006PASP..118.1711W}, \target's redshift of 0.015 corresponds to a luminosity distance of 67.5 Mpcs.

\subsection{NEOWISE IR light curves}
\begin{figure*}[ht]
    \centering
    \hspace{-0.1in}
   \includegraphics[width=0.6\textwidth]{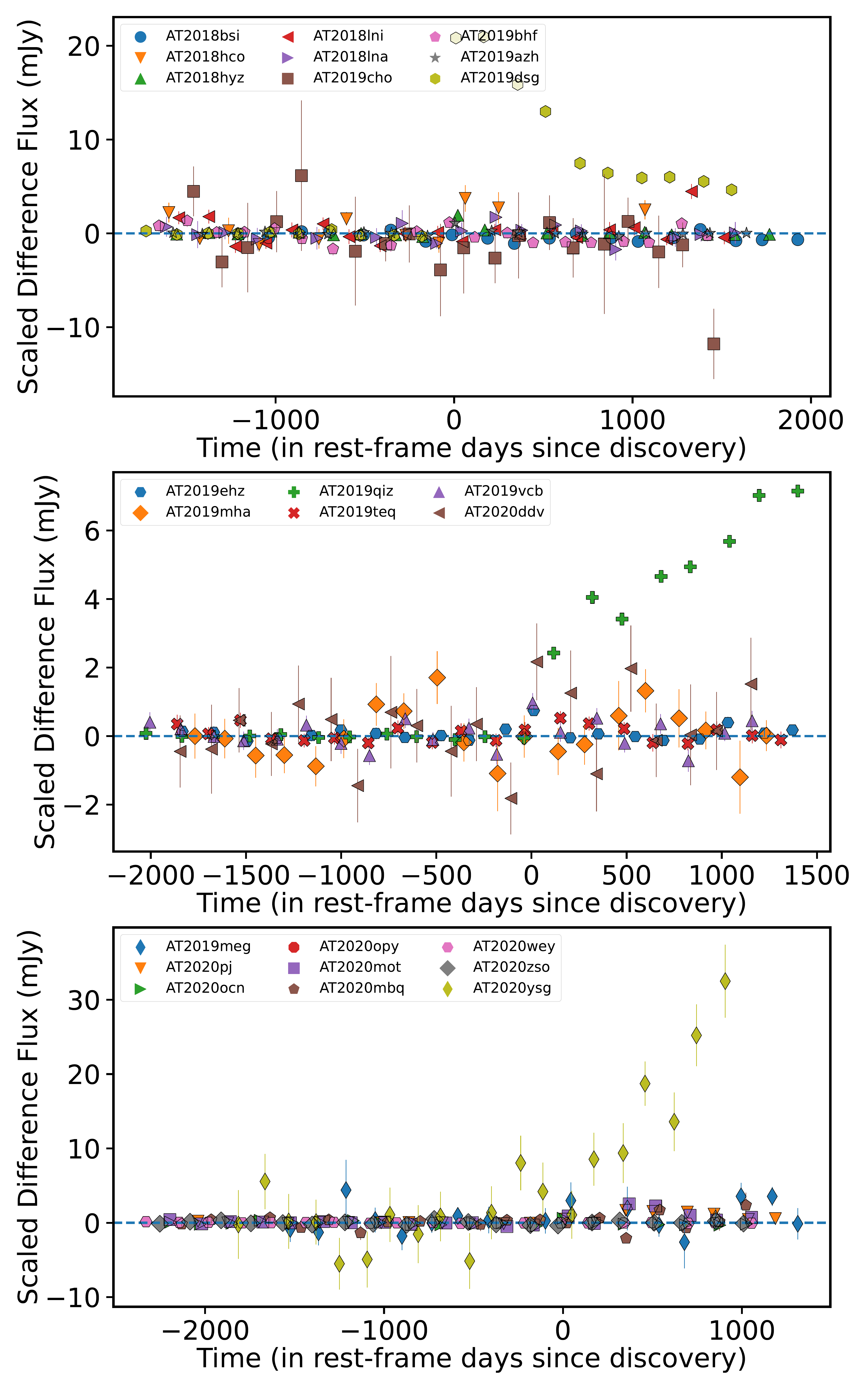}
   \caption{IR difference flux of ZTF-I TDEs scaled to \target's redshift of 0.015 vs time. The sources are distance-corrected by multiplying their observed IR difference flux with (luminosity distance of the TDE/luminosity distance of \target(=67.5 Mpcs))$^{2}$. Only two TDEs show a rising behavior: \target (a known QPE source) and AT2020ysg. We argue here that QPEs can explain the IR behavior and temperature (see Figure \ref{Fig:fig2}).}
   \label{Fig:fig1}
\end{figure*}

We extracted the IR light curves of all the ZTF TDEs (the ZTF-I sample) presented in \cite{2023ApJ...942....9H}. We obtained host-subtracted/difference IR magnitude/fluxes and co-added the observations to get a 6 month cadence. The data are shown in Fig.~\ref{Fig:fig1} and are divided into three panels for clarity (AT2018jbv, AT2020riz and AT2020qhs have large error bars consistent with no IR variations and 
are not shown, also for clarity). 

\subsection{Blackbody Fits}

Each NEOWISE observation of AT2019qiz that had an IR excess was fit with a blackbody model. 
These fits used Markov Chain Monte Carlo methods and a forward modeling approach, employing the WISE W1 and W2 filter profiles from the Spanish Virtual Observatories Filter Profile Service \citep{rodrigo12}. To keep our fits relatively unconstrained, we ran each of our blackbody fits with flat temperature priors of 100 K $\leq$ T $\leq$ 5000 K. The blackbody evolution is shown in Figure \ref{Fig:fig2}. 

The temporal evolution of the dust blackbody radius and temperature are broadly similar to other luminous IR dust echoes \citep[e.g.,][]{2021SSRv..217...63V, hinkle24}. A blackbody temperature that begins near or above the sublimation temperature for typical silicate grains and subsequently cools is consistent with previous studies on transient IR echoes \citep[e.g.,][]{jiang17, hinkle24}. Additionally, the late-time temperature plateau of $\approx$800 K is similar to other IR flares \citep[e.g.,][]{jiang21, hinkle24}. Conversely, the continually rising IR luminosity is uncommon for a TDE \citep{vanvelzen16, jiang21}, with most TDEs exhibiting short and relatively weak IR echoes.

\section{Model}
\label{sec:model}
\begin{figure*}[ht]
    \centering
    \hspace{-0.1in}
   \includegraphics[width=0.6\textwidth]{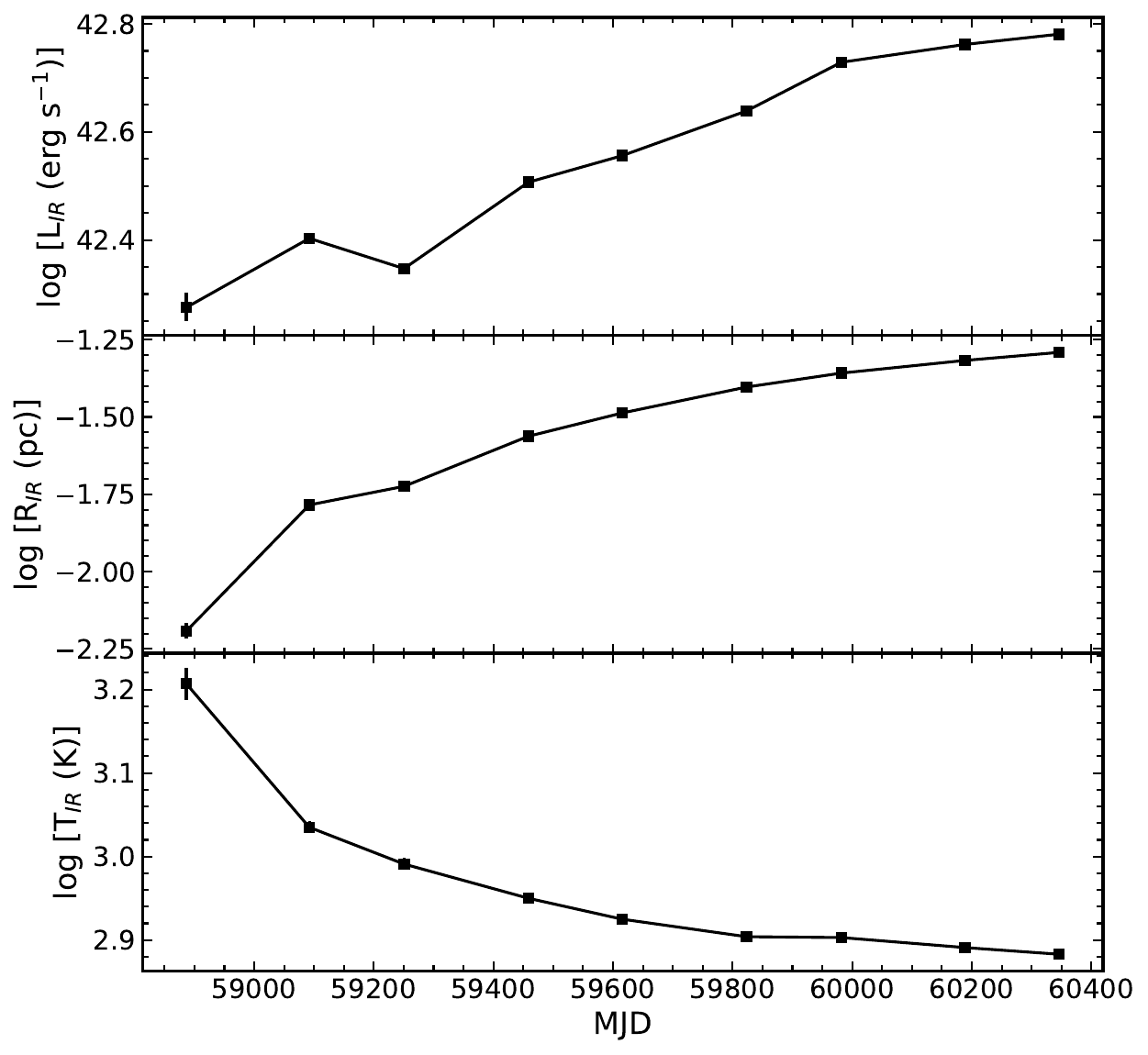}
   \caption{Temporal evolution of the IR blackbody luminosity (top panel), effective radius (middle panel), and temperature (bottom panel) for AT2019qiz.}
   \label{Fig:fig2}
\end{figure*}

The dust echo rises linearly and starts to plateau around day 1000, and the inferred temperature remains roughly constant at $\sim 1000$ K over that time (see the green crosses in the middle panel of Figures \ref{Fig:fig1} and the bottom panel of \ref{Fig:fig2}). The constancy of the temperature implies that, assuming the dust is re-radiating quasi-thermally after being irradiated by the central source, it is confined to a region of $\sim$ constant radius. With a rise time of $t_{\rm rise} = 1000$ days, the radius of the dust shell is 
\begin{equation}
    R_{\rm shell} = \frac{c t_{\rm rise}}{2} \simeq 0.4\textrm{ pc}.
\end{equation}
where the factor of 2 comes from the assumption that the shell is optically thin and $c \simeq 3\times 10^{10}$ cm s$^{-1}$ is the light speed.

\subsection{The TDE cannot produce a rising dust echo}
It is not possible to produce a dust echo that rises continuously over 1000 days by a TDE that rises and fades on a timescale of $\sim 30$ days (i.e., the optical/UV outburst associated with AT2019qiz; \citealt{2020MNRAS.499..482N}), 
because the peak temperature of the TDE-irradiated dust -- from which most of the reprocessed emission arises -- remains visible to the observer until $t_{\rm rise}$, implying that the reprocessed dust lightcurve would remain nearly constant over $\sim t_{\rm rise}$ before decaying rapidly. To see this explicitly, we model the luminosity of the TDE as
\begin{equation}
    L_{\rm tde}(t) = L_{\rm p}\frac{t/t_{\rm p}}{1+\left(t/t_{\rm p}\right)^{8/3}}, \label{Ltde}
\end{equation}
which is a simple phenomenological function that rises, peaks on a timescale of $\simeq 0.83\, t_{\rm p}$ at a value of $\simeq 0.52 \, L_{\rm p}$, and declines as $\propto t^{-5/3}$ (where $t$ is time). Letting the dust absorb incident radiation from the TDE and re-radiate as a blackbody, the instantaneous dust temperature is
\begin{equation}
    T(t) = T_{\rm d}\left(\frac{t/t_{\rm p}}{1+\left(t/t_{\rm p}\right)^{8/3}}\right)^{1/4}, \label{Toft}
\end{equation}
where
\begin{multline}
    T_{\rm d} = \left(\frac{L_{\rm p}}{4\pi \sigma R^2}\right)^{1/4}\times f \\
    \simeq 1000\textrm{ K}\left(\frac{L_{\rm p}}{10^{45}\textrm{ erg s}^{-1}}\right)^{1/4}\left(\frac{R}{0.4\textrm{ pc}}\right)^{-1/2} f. \label{Td}
\end{multline}
is $\sim$ the dust temperature near the peak of the TDE (more specifically our model thereof). There is a difference in the definition of $t = 0$ between Equations \eqref{Ltde} and \eqref{Toft}, as the latter only holds when the radiation from the central source has propagated to radius $R$, but this distinction is irrelevant for this simple model and henceforth 
we use the definition of $t = 0$ in Equation \eqref{Toft}. 
Equation \eqref{Toft} also ignores the efficiencies of absorption and emission and any sublimation, and assumes the dust is geometrically in the form of a spherical shell; we incorporate these and others uncertainties in the parameter $f \le 1$. The sublimation timescale is $\gg 1000$ days for graphite grains at these temperatures (e.g., \citealt{guhathakurta89, waxman00, lu16, vanvelzen16}).

The emitted luminosity is (at least for a pure blackbody) equal to the incident luminosity, but the observed/received luminosity is complicated by retarded time effects: the radiation from a patch of the shell inclined by an angle $\theta$ with respect to the axis that points to the observer is received at a time
\begin{equation}
    t_{\rm r}(\theta) = t-\frac{R}{c}\left(1-\cos\theta\right).
\end{equation}
The received luminosity is an integral over the sphere at the delay time appropriate to each angle $\theta$:
\begin{equation}
    L_{\rm tde, rec} = \frac{\tilde{f}}{2}L_{\rm p}\times \int_{0}^{\theta_{\rm max}(t)}\left(\frac{T(t_{\rm r})}{T_{\rm p}}\right)^4\sin\theta d\theta,
\end{equation}
where
\begin{equation}
    \theta_{\rm max}(t) = \begin{cases}
        \arccos\left(1-\frac{ct}{R}\right) \quad & \textrm{ for }0\le t<2R/c \\
        \pi & \textrm{ for }t\ge 2R/c,
    \end{cases}
\end{equation}
which implies that we only ``see'' the entire shell after a time of $2R/c$. The parameter $\tilde{f}$, as for $f$, crudely accounts for uncertainties related to, e.g., the covering fraction of the dust and finite optical depth effects, and the factor of 2 ensures that $L_{\rm tde, rec} = L_{\rm p}$ when $\tilde{f} = 1$ and $R \rightarrow 0$.

\begin{figure}
    \includegraphics[width=0.495\textwidth]{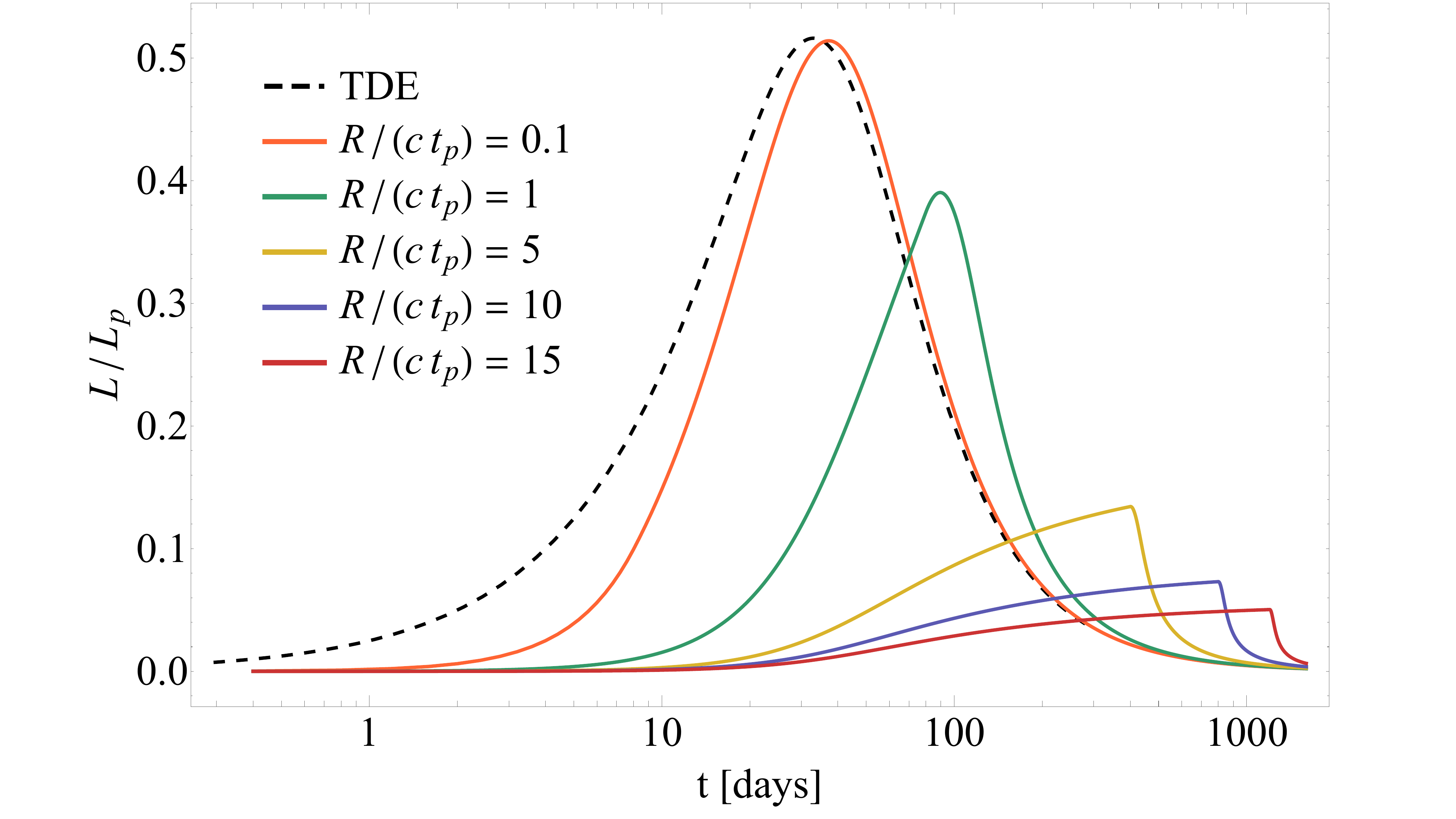}
    \caption{The luminosity of the dust echo caused by the TDE, which is shown by the black-dashed line. The colored curves place the dust shell at the radii given in the legend, where here $t_{\rm p} = 40$ days to give a TDE lightcurve that peaks on a timescale of $\sim 30$ days. When $R \gg c t_{\rm p}$, the reprocessed emission remains roughly constant until the light-crossing time of the shell, after which it rapidly fades.}
    \label{fig:L_Lpeak}
\end{figure}

Figure \ref{fig:L_Lpeak} illustrates the received luminosity from a TDE with $t_{\rm p} = 40$ days, such that the luminosity -- as shown by the black-dashed curve -- peaks on a timescale of $\sim 30$ days (recall that our phenomenological function $L_{\rm tde}$ peaks at $\simeq 0.83 t_{\rm p}$). The solid curves show the received luminosity from shells located at the radii given in the legend: when $R < c t_{\rm p}$, finite light-crossing time effects are small, and the received luminosity is nearly equal to the emitted luminosity. Alternatively, when $R \gg c t_{\rm p}$, the lightcurve reaches a plateau on a timescale of $\sim 2-3 t_{\rm p}$ before plummeting after the light-crossing time is reached. In summary, the IR light curves expected from TDEs (Fig.~\ref{fig:L_Lpeak}) are inconsistent with the linear rise seen in Fig. \ref{Fig:fig1}.

\subsection{The remnant TDE disk cannot explain the observed IR temperature}
Reprocessed TDE emission alone therefore cannot generate the IR lightcurve from AT2019qiz, but the linearly rising behavior of the dust echo can be straightforwardly reproduced by a constant-luminosity source that simply ``turns on'' at some time. There are two possible origins for this source: the TDE disc and the QPEs \citep{2019qiz}. The former has a (model-dependent) luminosity of $L_{\rm disc} \sim 10^{43}$ erg s$^{-1}$. The latter has a comparable average luminosity -- each eruption releases an energy of $E_{\rm qpe} \simeq 10^{48}$ erg, which yields a mean luminosity of $\bar{L}_{\rm qpe} = E_{\rm qpe}/(2\textrm{ days}) \simeq 10^{43}$ erg s$^{-1}$ over the recurrence time of $2$ days. With an observing cadence $\gtrsim 2$ days, each of these sources would appear to produce a linearly rising dust echo that plateaus on a timescale of $t_{\rm rise}$, after which it would remain constant. 

\begin{figure}
    \includegraphics[width=0.495\textwidth]{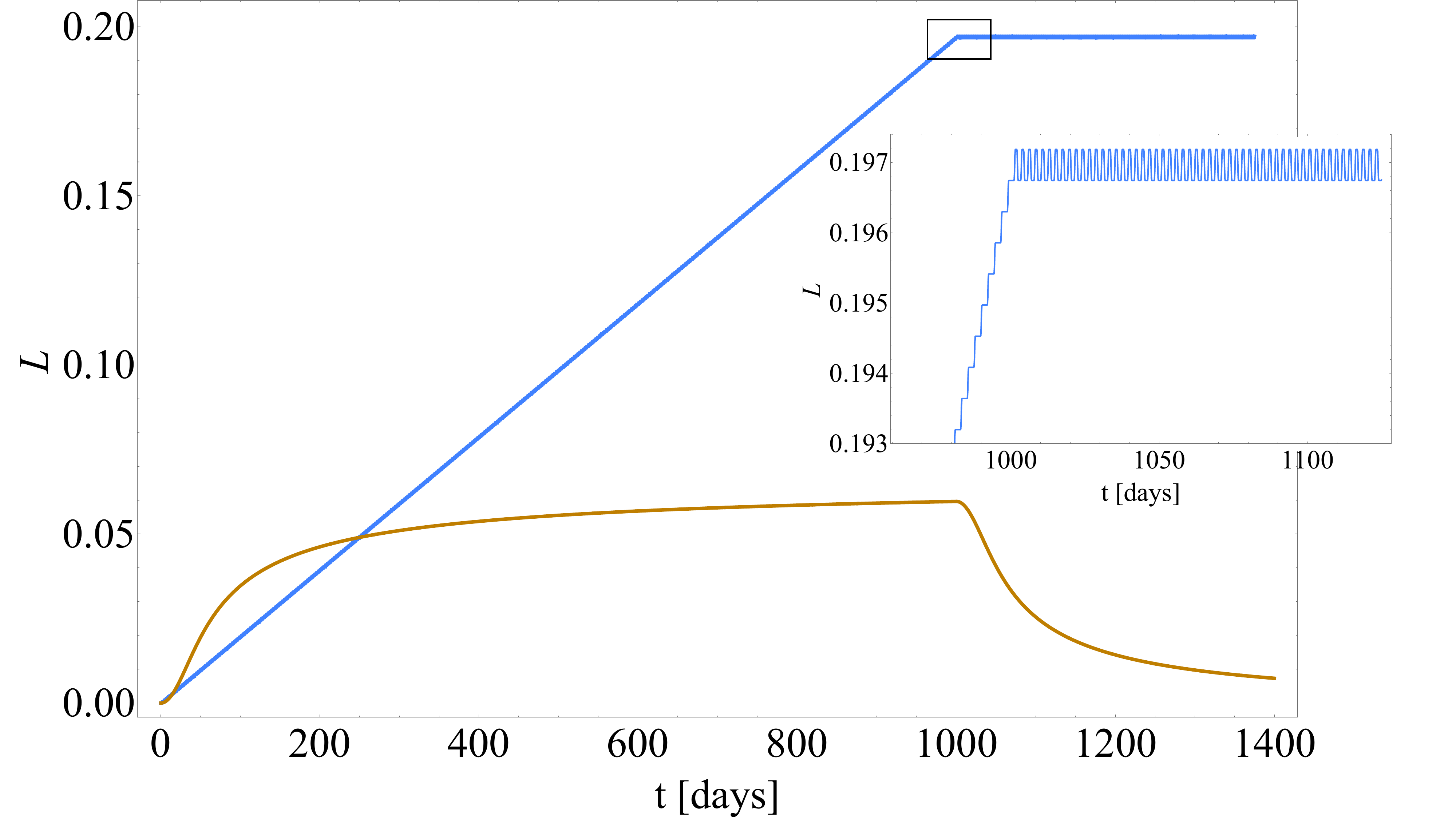}
    \caption{The luminosity of the dust echo caused by a TDE (brown curve) with $R/(ct_{\rm p}) = 12.5$, or -- with $t_{\rm p} = 40$ days -- $R = 0.4$ pc, and the QPEs (blue curve). The latter appears to consist of a linear rise over 1000 days followed by a subsequent plateau. However, the inset demonstrates that the linear rise actually consists of step-like increases that occur on the two-day recurrence time of the QPEs, while the flat phase consists of small-amplitude oscillations on the same recurrence time.}
    \label{fig:qpe_luminosity_w_TDE}
\end{figure}

There are, however, two differences between them. First, if we model the QPEs as a sum of 2-day-separated outbursts that rise and fade over $\sim 8$ hours, then the dust echo will rise in step-like increments that plateau on a $\sim 8$ hour timescale, analogous to the reprocessed emission from the TDE when $R \gg c t_{\rm p}$ (i.e., because the rise time is much less than the light-crossing time of the shell). After the light-crossing time of the dust shell, the dust echo will be \emph{nearly} constant, but the relative amplitude will ``flicker'' on the recurrence time of the QPEs by an amount $\sim 2\textrm{ days}/(1000\textrm{ days} \simeq 0.2\%$ (i.e., the ratio of the recurrence time to the light-crossing time) as new QPE outbursts reach the dust shell and the peaks of past QPEs fade. To demonstrate this behavior explicitly, the blue curve in Figure \ref{fig:qpe_luminosity_w_TDE} shows the reprocessed dust emission from a dust shell with light-crossing time of $1000$ days (or radius $R \simeq 0.4$ pc) and a model in which the QPE luminosity is a sequence of Gaussian outbursts, such that the outburst peaks are separated by two days and the width of each outburst is $\sim 8$ hours. The inset shows that -- while the lightcurve seems to rise linearly to a constant when averaged over timescales $\gtrsim 2$ days -- the increase in luminosity is composed of step-like increases on the recurrence time of 2 days, followed by oscillations on the same recurrence time at the $\sim 0.1\%$ level as past outbursts fade and new outbursts irradiate the dust shell. The brown curve gives the corresponding emission from the TDE (still assuming $t_{\rm p} = 40$ days, i.e., the black-dashed curve in Figure \ref{fig:L_Lpeak}), which reinforces the conclusion that the linearly rising trend exhibited by AT2019qiz cannot be reproduced by the TDE alone. The peak of each individual outburst in the QPE model was set to $2\times L_{\rm p}$, which is mostly for clarity in representing both lightcurves alongside the inset, but we note that the peak bolometric luminosity inferred from the TDE in \target of $\sim 10^{43.6}$ erg s$^{-1}$ \citep{2020MNRAS.499..482N} is -- qualitatively consistent with this normalization -- a factor of a few smaller than the individual QPE luminosity. 

The second difference between the two mechanisms is the dust temperature: the peak temperature associated with the reprocessed emission (still maintaining that the dust absorbs incident emission and re-radiates as a blackbody) is related to the instantaneous luminosity via $T \propto L^{1/4}$. If the dust were irradiated only by the disc, which has a constant luminosity of $10^{43}$ erg s$^{-1}$, the temperature would be a factor of $\sim 3$ below that produced by the luminosity of the TDE itself, or $\sim 300$ K. This temperature is significantly below that inferred from modeling the observed flux (between $\sim 800-1200$ K; see Figure \ref{Fig:fig2}). Alternatively, the peak bolometric luminosity of the QPEs is estimated to be $\sim 10^{44-45}$ erg s$^{-1}$, such that the corresponding dust temperature would be in better agreement with what is observed. 


\section{Summary, Conclusions, and Predictions}\label{sec:summ}
The linearly increasing IR flux over $\sim 1000$ days from AT2019qiz is (nearly) unprecedented among known TDEs, and from this behavior our main conclusions are the following: 
\begin{enumerate}
    \item{The emission from the TDE itself, which rises and fades over $\sim 30$ days, cannot generate the dust echo from \target, which rises roughly linearly over $\sim 1000$ days.}
    \item{In principle, both the late-time/plateau emission from the TDE disc and the recurrence-time-averaged QPE emission can reproduce the qualitative behavior of the dust echo, provided that the dust is in the form of a shell with a radius of $\sim 0.4$ pc.}
    \item{However, the dust temperature of $\sim 800-1200$ K, as inferred from the WISE W1 and W2 band observations, cannot be reached by a constant disc luminosity of $10^{43}$ erg s$^{-1}$, which would instead yield a temperature a factor of $\sim 3$ smaller; the instantaneous luminosity associated with the QPEs -- being comparable to that of the original TDE -- can generate dust temperatures that are consistent with observations.}
    \item{The final, constant-luminosity dust echo is then equal to\footnote{Strictly speaking, the flux is detected in the WISE bands and then converted to an intrinsic luminosity; the disc luminosity should also contribute to the emission, but at a significantly lower temperature.} the recurrence-time-averaged luminosity of the QPEs, modulo the dust covering fraction; this result is consistent with the blackbody fitting results shown in Figure \ref{Fig:fig2}, provided that the covering fraction is high ($\gtrsim 50\%$ for a recurrence-time-averaged QPE luminosity of $\sim 10^{43}$ erg s$^{-1}$).}
\end{enumerate}
Taken together, these four points suggest that the QPEs associated with the TDE AT2019qiz are responsible for its unique dust echo.

There are two predictions of this model that render it falsifiable:
\begin{enumerate}
    \item{The IR luminosity from the QPEs should be $\sim$ constant on timescales longer than their 2-day recurrence time, which is consistent with observations, but should also display periodic variations (i.e., flickering) at the $\sim 0.1\%$ level on this same timescale, which arise as new QPEs illuminate the dust shell and past QPEs fade after they traverse the entire shell on the light-crossing time. High-resolution imaging (high precision photometry) by, e.g., JWST, could detect such behavior, and the non-detection of this flickering would strongly rule against this interpretation.}
    \item{The QPEs from other sources, including GSN 069 \citep{2023A&A...674L...1M}, eRO-QPE1 \citep{2024ApJ...965...12C, aliveandbare}, and AT2019vcb \citep{bykov}, dim and occasionally disappear over longer timescales ($\sim$ years). If the QPEs from AT2019qiz exhibit the same secular trend as these other sources, we would expect the dimming to be reflected in the IR emission, and in the extreme limit of a sudden and instantaneous cessation of QPEs the IR emission would fade linearly over the light-crossing time. Conversely, if the IR emission displays such a linear decline in flux, it would suggest that the QPEs from AT2019qiz have shut off. The lack of such correlations would disfavor this model.}
\end{enumerate}

Finally, we note that the dust echo from the TDE AT2020ysg (z=0.277; luminosity distance of 1467 Mpc) displays a similar, linear rise in IR flux over comparable timescales to \target. In fact, at 1000 days (rest frame), AT2020ysg's IR flux (corrected for redshift) is a factor of 5 higher than \target. Consequently, we identify AT2020ysg as a potential QPE source which should be investigated with high-cadence X-ray follow-up. \target's peak QPE soft X-ray (0.3-2.0 keV) flux is $\sim$10$^{-11}$ erg/s/cm$^{2}$. If we assume AT2020ysg's putative QPEs are similar in strength to \target then the we would require a soft X-ray sensitivity of $\sim$ 10$^{-11}$ erg/s/cm$^{2}$ $\times$ (67.5 Mpc/1467 Mpc)$^{2}$ $\sim$ 2$\times$10$^{-14}$ erg/s/cm$^{2}$ at QPE peak, which can currently be achieved with \xmm and \chandra. 

\subsection{Caveats}
A caveat related to the first of these two points is related to our assumption about the geometry of the dust: we assumed that the dust is in the form of a smooth and spherically symmetric shell, but obviously and realistically there will be smaller-scale fluctuations in its distribution in the circumnuclear medium. These fluctuations will be imprinted on the reprocessed emission and lead to variability, but this additional variability should itself repeat on the recurrence time of the QPEs. A Fourier analysis should, therefore, still yield excess power at the frequency associated with the QPE recurrence time, even if the time-domain lightcurve is considerably noisier due to geometrical effects. 

In addition to geometry, there are a number of other inherent uncertainties that result from our simplistic model, such as finite-opacity effects, and these could explain the more gradual flattening of the lightcurve seen in Figure \ref{Fig:fig2} (i.e., if the dust is optically thick at some distance through the shell, the difference in the projected geometry results in a lightcurve that reaches its plateau quadratically with time instead of linearly). We also assumed that the dust is at a single radius from the black hole, which is consistent with the fact that the dust temperature does not evolve significantly, but realistic and finite-volume effects would both lead to differences in the temperature evolution (and dust at smaller radii could explain the higher temperatures inferred from the WISE modeling; see Figure \ref{Fig:fig2}) and the spectrum itself. Finally, dust sublimation could also lead to additional evolution of the effective temperature and luminosity, and while the temperatures inferred for \target are substantially below the sublimation temperature for carbonaceous grains, such that the corresponding effects of sublimation for \target are small, the time-dependent removal of dust could be more important for higher-temperature echoes (or, equivalently, dust at smaller distances from the black hole).


\subsection{Predictions for the Roman survey}
Infrared surveys can discover slowly rising lightcurves, thus providing a promising avenue for identifying potential QPEs for follow-up with X-ray surveys. We focus here on the Nancy Grace Roman Space Telescope, which will conduct a time-domain survey over a period of 2 years, optimized for finding high-redshift type Ia supernovae \citep{2021arXiv211103081R}. Two surveys with a 5 day cadence are planned: a wide survey (19 deg$^{2}$ in the RZYJ filters) and a deep survey (4.2 deg$^{2}$) with a longer exposure time and using a redder set of filters (YJHF). 

To estimate the expected number of QPE-driven echoes we use the \target's properties after 1 year: an IR blackbody luminosity of $10^{42.5}\,{\rm erg}\,{\rm s}^{-1}$. For a given dust temperature, we can then compute the observed magnitude for the Roman filters at each redshift. To find the maximum redshift we have to adopt a flux limit. Because the dust echoes are slowly evolving, we can co-add four Roman images (reducing the cadence to 20 days) to reach one magnitude deeper than the single-epoch limits listed in \citet{2021arXiv211103081R}. Given the relatively cool temperature of the dust ($\sim 10^3$~K
), the Roman filters are on the Wien tail of the spectrum and the reddest filter will be most useful to detect the echoes.  We thus use the reddest filter for the Roman wide/deep survey to find the maximum redshift and the volume that can be probed by each survey. We then multiple this volume with a QPE rate of $0.6_{-0.4}^{+4.7}
,{\rm Mpc}^{-3}{\rm yr}{-1}$ \citep{2024A&A...684L..14A} and make the simplifying assumption that 50\% of X-ray emitting QPEs also produce dust echoes similar to \target.  In Table~\ref{tab:rates} we show the results for two different dust temperatures consistent with the sample of nuclear IR flares reported by \citet{2024arXiv240701039N}. We see that unless the dust is always very cold we can expect a sizable number of QPE-powered echoes in both Roman time domain surveys. 

\begin{table}[]
    \centering
    \begin{tabular}{c|c c}
        Dust temperature & Wide Survey & Deep Survey \\
        \hline
        $T=1000$~K & 0.04-1 & 0.3-8\\
        $T=1800$~K & 1-32 & 2-60 
    \end{tabular}
    \caption{Number of QPE-powered dust echo detections in the two Roman time domain surveys described in \citet{2021arXiv211103081R} using the QPE volumetric rate of \citet{2024A&A...684L..14A}, and assuming 50\% of X-ray QPEs are ``bright'' (bolometric luminosities $\gtrsim$10$^{44}$ erg/s) and have host galaxies with dust similar to \target. For each combination of survey and the dust temperature, the range in the detection rate follows from the statistical uncertainty on the QPE rate.}
    \label{tab:rates}
\end{table}

Finally, we checked the NEOWISE IR data of the other QPEs known in the literature -- eRO-QPE1 \citep{2024ApJ...965...12C, aliveandbare}, eRO-QPE2 \citep{2024A&A...690A..80A, 2024arXiv241100289P}, eRO-QPE3 \citep{2024A&A...684A..64A}, eRO-QPE4 \citep{2024A&A...684A..64A}, and RX1301 \citep{2024A&A...692A..15G}. None of them show a rising IR light curve as seen in \target, which can be attributed to their comparatively faint luminosities: $L_{\rm QPE1} \lesssim few\times 10^{43}$ erg s$^{-1}$, $L_{\rm QPE2} \lesssim 10^{42}$ erg s$^{-1}$, $L_{\rm QPE3} \lesssim few\times 10^{41}$ erg s$^{-1}$, $L_{\rm QPE4} \lesssim 10^{43}$ erg s$^{-1}$, and $L_{\rm RX1301} \lesssim few\times 10^{41}$ erg s$^{-1}$. The corresponding recurrence-time-averaged luminosities of each source would also be less than that for \target, making both the temperatures and overall energetics of these other QPE systems much lower, and hence the non-detections are consistent with the model proposed here. 
\newline




\noindent{}D.R.P. would like to thank  
Amit Kumar Mandal for preliminary IR light curve extractions which inspired him to dig deeper into this work. D.R.P also thanks Thomas Wevers for fruitful discussion about this work that help improve it. D.R.P ~was partly funded by a NASA/Swift grant for this work (award number 035196-00001). E.R.C.~acknowledges support from NASA through the Astrophysics Theory Program, grant 80NSSC24K0897.

\newpage
\bibliographystyle{aasjournal}
\bibliography{ms}

\end{document}